\documentclass[preprint,12pt]{article}
 
\usepackage{amssymb}
\usepackage{caption}
\usepackage{breqn}
\usepackage{graphicx}

\begin{document}

\textbf{Hermitian Separability of BFKL eigenvalue in  Bethe–Salpeter approach}
\begin{center}

\small
Mohammad Joubat$^{(a)}$   and Alex Prygarin$^{(b)}$  
\\
$^{(a)}$ Department of Mathematics, Ariel University, Ariel 40700, Israel\\
$^{(b)}$ Department of Physics, Ariel University, Ariel 40700, Israel 
\end{center}
\normalsize

\normalsize

\begin{abstract}
   We consider the Bethe–Salpeter approach to the BFKL evolution 
   in order to naturally incorporate the property of the Hermitian Separability in the BFKL approach. We combine the resulting all order ansatz for the BFKL eigenvalue together with reflection identities for harmonic sums and derive the most complicated term of the next-to-next-to-leading order BFKL eigenvalue in SUSY $N=4$. We also suggest  a numerical technique for reconstructing  the unknown functions in our ansatz from the known results for specific   values
   of confomal spin.

\end{abstract}

\newpage
\section{Bethe-Salpeter approach to BFKL equation }\label{}
The Balitsky-Fadin-Kuraev-Lipatov~(BFKL)~\cite{BFKL} equation is traditionally schematically written in the form of the linear Schr{\"o}dinger equation 
\begin{eqnarray}\label{scroedinger}
H \psi = E \psi
\end{eqnarray}
for the BFKL Hamiltonian $H$ and the BFKL  eigenvalue  $E$ which is related to the  pomeron intercept. The eigenfunction $\psi$ is a complex function of two transverse degrees of freedom, either the transverse two dimensional  momentum or its canonic  conjugate, the transverse coordinates. 
The BFKL eigenvalue depends on two  real valued  degrees of freedom, the anomalous dimension $\nu$ and the conformal spin $n$ emerging through 
Mellin transform of the two-dimensional transverse momentum. For the singlet BFKL equation the BFKL eigenvalue is a function of the complex variable~\footnote{In this paper we follow notation of N.~Gromov, F.~Levkovich-Maslyuk and G.~Sizov~\cite{GROMOV} and  use $\nu$ divided by two, instead of the traditional notation   of  anomalous dimension and conformal spin through  $i\nu +\frac{|n|}{2}$.} 
\begin{eqnarray}
z=-\frac{1}{2}+ \frac{i\nu}{2} +\frac{|n|}{2}
\end{eqnarray} 
for continuous $\nu$ ranging from $-\infty$ to $\infty$ and discrete
$n=0, \pm 1, \pm 2, ...$.  

The analytic expressions for the BFKL eigenvalue in the color singlet channel are currently available only for the leading order~(LO) and 
next-to-leading order~(NLO)~\cite{NLO,Kotikov:2000pm,Kotikov:2001sc,Kotikov:2002ab}   of the perturbative expansion in both QCD and $N=4$ SYM theory.  There is also some information available for next-to-next-to-leading order~(NNLO) in the  $N=4$ SYM, which follows from modern integrability techniques~\footnote{See  recent review paper discussing different aspects of integrability techniques applied to the BFKL evolution~\cite{Alfimov:2020obh}}. The information currently available about the NNLO BFKL eigenvalue is analytic expressions as functions of $\nu$ for any particular integer value of $n$ in $N=4$ SYM as well as analytic expressions for arbitrary integer values of $n$ for a case of $\nu=0$ also in $N=4$ SYM. Not that much known about NNLO BFKL eigenvalue in QCD  except  for $\nu \to \infty$ limit. For the most updated information regarding the known information on the NNLO BFKL eigenvalue  the reader is refereed to Ref.\cite{Alfimov:2020obh}. We discuss this in more details in the next section. 

The NLO eigenvalue is expressed through more complicated functions compared to those present at the LO level.  In the present paper we focus on one major feature of  the LO and the NLO functions, namely the so-called Hermitian separability first discussed by A.~Kotikov and L.~Lipatov~\cite{Kotikov:2001sc, Kotikov:2002ab}. By Hermitian separability one  means a possibility of writing a function of complex variable $z$ and its complex conjugate $\bar{z}$ as a sum 
two contributions separately dependent on $z$ and $\bar{z}$ 
\begin{eqnarray}\label{hermit}
f(z, \bar{z})= F(z)+F(\bar{z})
\end{eqnarray}
In our case we restrict $F$  to be the same function for $z$ and $\bar{z}$, which reduces any related calculations to much simpler one dimensional problem of computing only one function. In the case of the BFKL eigenvalue the function $F(z)$ is the    single valued function of a complex variable so that $f(z, \bar{z})$ is always real for any value of $z$. 
 
\begin{figure}[h!]
  \includegraphics[width=0.7\linewidth]{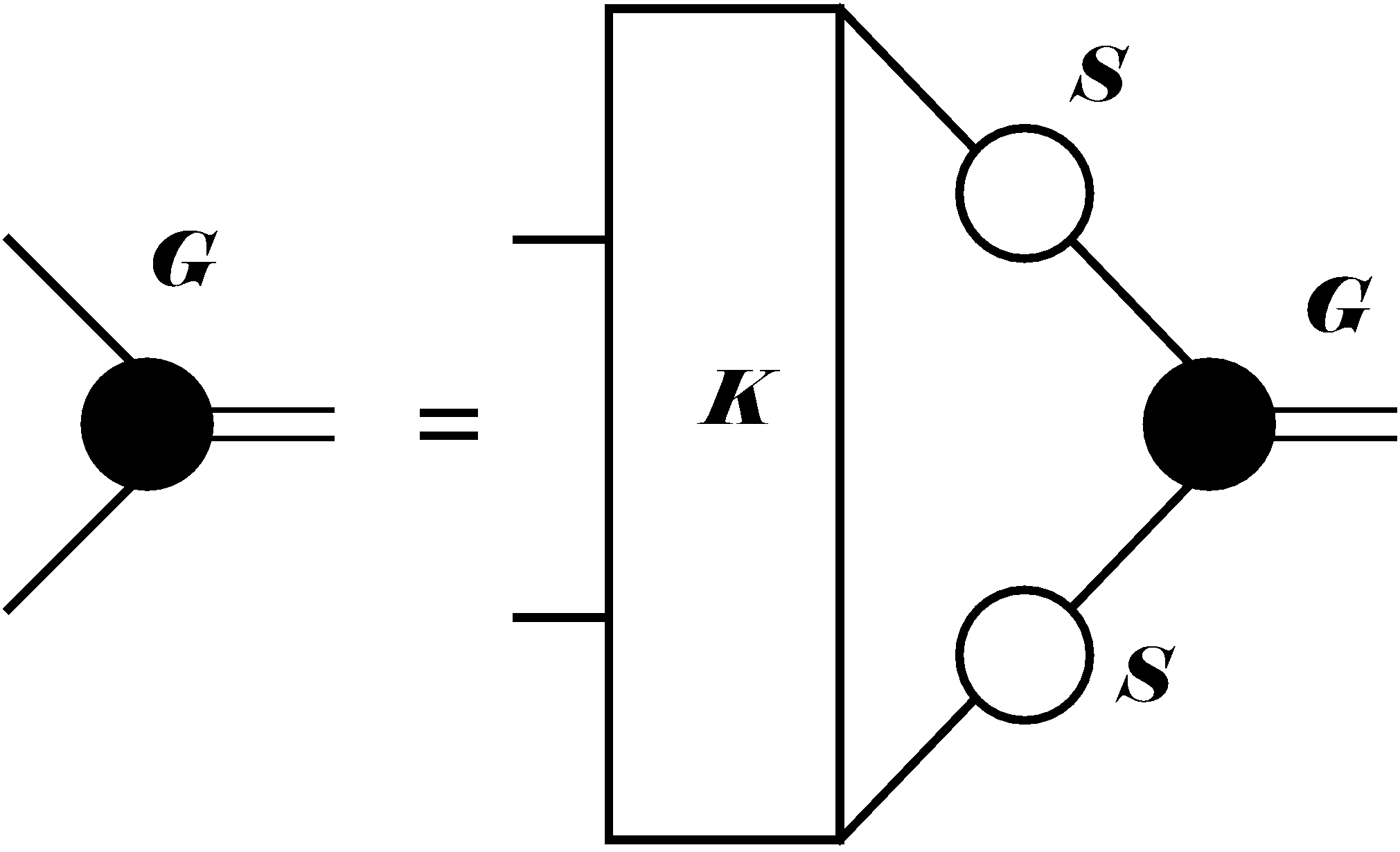}
  \caption{The figure shows graphical representation of the Bethe-Salpeter equation in eq.~\ref{BES}.    }
  \label{fig:BES}
\end{figure}

 The LO eigenvalue is manifestly Hermitian separable, whereas the NLO eigenvalue~\cite{NLO} is not. It was demonstrated  by A.~Kotikov and L.~Lipatov~\cite{Kotikov:2001sc, Kotikov:2002ab} that color singlet  NLO eigenvalue in $N=4$ SYM  can be written as 
 a combination of a product of two hermitian separable functions and
a hermitian separable function
 \begin{eqnarray}\label{NLO}
f^{NLO}(z,\bar{z})=f^{LO}(z,\bar{z})g(z,\bar{z})+\rho(z, \bar{z})
\end{eqnarray} 
where $f^{LO}(z,\bar{z})$ is the corresponding LO eigenvalue. The function $f^{LO}$, $f^{NLO}$, $g$ and $\rho$ all have Hermitian separable form of eq.~(\ref{hermit}).
 This sort of non-linearity is difficult to explain by the Schr{\"o}dinger equation approach to the BFKL dynamics, where the two degrees of freedom corresponding to the complex variables $z$ and $\bar{z}$ are mixed   at the Hamiltonian level.   Based on the works of A.~Kotikov and L.~Lipatov~\cite{Kotikov:2001sc,Kotikov:2002ab}  it is natural  to consider another evolution  where the two degrees of freedom are separated already at the level of the kernel of the corresponding  equation. 
A  natural choice for describing a bound state of two reggeized gluons would be to use the Bethe-Salpeter equation, which was originally constructed to describe  quantum bound states~\cite{BSE} and having a variety of applications in quantum field theory (positronium, mesons etc.). The Bethe-Salpeter equation can be schematically represented as  follows
\begin{eqnarray}\label{BES}
G= K \otimes S \otimes G \otimes S
\end{eqnarray}
where $G$ is the propagator of the bound state under consideration, $K$ is the kernel and $S$ is the bare propagator (see Fig.~\ref{fig:BES}). 

In the Bethe-Salpeter approach  one can represent the BFKL dynamics as 
pole decomposition of the scattering amplitude  in the plane of complex angular momentum $j$. 
The leading singularity of $j \to 1$ corresponds to the Regge kinematics in which the original BFKL was derived. 
It is customary to denote $j=1+\omega$ and make an expansion  in powers of $\omega$. The leading-order~(LO) contributions would correspond to the simple pole $1/\omega$ of the BFKL amplitude (the propagator of the bound state of two reggeized gluons), the next-to-leading~(NLO) contributions would also include a free term $(\omega)^0$, the next-to-next-to-leading~(NNLO) would account for the first order in $\omega$ and so on. This happens because of  the integrability property of the BFKL equation, which allows to get rid of the integration of the transform degrees of freedom and write the final expression as the product of the eigenvalue times the eigenstates. 
 Due to recursive structure of the Bethe-Salpeter equation in eq.~(\ref{BES}) the  sum of all those contribution should   equal $\omega$ itself.
 This can be written as follows~\cite{BARTELS} 
\begin{eqnarray}\label{Bethe}
1= \frac{a}{\omega} \sum^{\infty}_{i=0} \omega^i
 \sum_{k=0}^{\infty} a^{k} f_{i,k},
\end{eqnarray}
where $a=\frac{\alpha_s N_C}{2\pi}$ is the coupling constant. 
We assume  that the functions 
$f_{i,k}$ are Hermitian  separable at any order and can reproduce the
structure of the next-to-leading eigenvalue and make prediction for the  next-to-next-to-leading eigenvalue  in the following  way.
Let us denote the leading order eigenvalue by 
\begin{eqnarray}
\omega_{0}=a f^{LO}(z,\bar{z}),
\end{eqnarray}
the next-to-leading order eigenvalue by 
\begin{eqnarray}
\omega_{1}=a^2 f^{NLO}(z,\bar{z})
\end{eqnarray}
and so forth. 
The leading order BFKL eigenvalue  in $N=4$ SYM is rather simple 
\begin{eqnarray}\label{flo}
f^{LO}(z,\bar{z})=4\left( -\psi( z+1)-\psi(\bar{z}+1) +2 \psi(1) \right)
\end{eqnarray} 
and the corresponding NLO expression reads~\cite{Kotikov:2001sc,Kotikov:2002ab}
\begin{eqnarray}\label{fnlo}
 f^{NLO}(z,\bar{z})&=& \Phi(z+1) +\Phi(\bar{z}+1)\\
 &&-\frac{1}{2} f^{LO}(z,\bar{z}) \left(\beta'(z+1)+\beta'(\bar{z}+1)+\frac{\pi^2}{6}\right), \;\;  \nonumber
\end{eqnarray}
where the functions $\Phi(z)$ and $\beta'(z)$ are given by 
\begin{eqnarray}\label{beta}
\beta'(z)=\sum_{r=0}^{\infty} \frac{(-1)^{r+1}}{(z+r)^2}
\end{eqnarray}
and 
\begin{eqnarray}\label{phi}
\Phi(z)= 3 \zeta(3) +\psi^{''} (z) +2 \Phi_2(z)+2 \beta'(z) \left(\psi(1)-\psi(z)\right). 
\end{eqnarray}
The function $\Phi_2(z)$ is the most complicated function and it is defined as follows 
\begin{eqnarray}
\Phi_2(z)= \sum_{k=0}^{\infty} \frac{\beta'(k+1)+ (-1)^k \psi'(k+1)}{k+z}-\sum_{k=0}^{\infty}\frac{(-1)^k (\psi(k+1)-\psi(1))}{(k+z)^2} \;\;\;
\end{eqnarray} 
In  section~\ref{comparison} we write these expressions in terms of harmonic sums (see eq.~(\ref{flos}) and eq.~(\ref{fnlos})).

To the required  next-to-next-to-leading~(NNLO) order eq.~(\ref{Bethe}) reads
\begin{eqnarray}
1=\frac{ a(f_{0,0}+a f_{0,1}+a^2 f_{0,2})}{\omega}
+
a (f_{1,0}+a f_{1,1})+a\omega f_{2,0}
\end{eqnarray}
plugging
 \begin{eqnarray}
 \omega = a (f^{LO}+a f^{NLO}+a^2 f^{NNLO}+..)
 \end{eqnarray}
 and expanding in the powers of the coupling constant we obtain the first three orders in the perturbation theory as follows. The LO eigenvalue
 \begin{eqnarray}
 \omega_0=a f_{0,0}
 \end{eqnarray}
 the NLO eigenvalue of order  $a^2$
\begin{eqnarray}\label{omega1}
 \omega_1=a \left(\omega_0 f_{1,0}+a f_{0,1} \right)
 =
 a^2 \left(f_{0,0} f_{1,0}+f_{0,1}\right)
\end{eqnarray}

and finally the NNLO eigenvalue of order  $a^3$.

\begin{eqnarray}\label{omega2}
 \omega_2 &=&a \left(\omega_0^2 f_{2,0}+a \omega_0 f_{1,1}+\omega_{1} f_{1,0}+
 a^2 f_{0,2} \right) \\
 &=&  a^3 \left( f^2_{0,0} f_{2,0}+f_{0,0}f_{1,1}+f_{0,0}f^2_{1,0}+
 f_{1,0}f_{0,1}+f_{0,2}\right)  \nonumber 
\end{eqnarray}

The  eq.~(\ref{omega2}) is our master equation for the NNLO BFKL eigenvalue, which we  discuss in more details below. In the next two sections we focus on possible to ways to apply it to the known NNLO results in $N=4$ SYM.

Our analysis shows that at the NNLO level the function $\omega_2$ is expressed in terms of  three unknown functions $f_{1,1}$, $f_{0,2}$  and $f_{2,0}$. The functions $f_{0,0}$, $f_{0,1}$ and  $f_{1,0}$ are known from the previous orders.  The functions $f_{i,j}$ are  single valued meromorphic  functions of different level of complexity.  The most complicated is
 $f_{0,2}$ function and the simplest is $f_{0,0}$. This can be shown by the following arguments. All currently available results show that the  BFKL eigenvalue is built of polygamma functions and its generalizations. Those functions are either logarithmically divergent at infinite value of the argument or give  transcendent constants~\footnote{Here we treat   Riemann zeta function at integer value as transcendent constant, despite the fact that even $\zeta(3)$ is still not proven to be transcendent constant.}. The transcendentality of 
 constant determines the "transcendentality" of the underlying function.   This concept despite  not being  rigorously  proven is very useful and widely used in building  functional bases for different ans{\"a}tze.  Another observation, which is also widely used is that    the maximal transcendentality is increased by two units 
 for each order of the perturbations theory. In this notation the function building  $\omega_0$, i.e. digamma function is assigned transcendentality one, the function building  $\omega_1$ are assigned 
 maximal transcendentality three and the functions building   $\omega_2$ all have maximal transcendentality five. The transcendentality is additive as functions are multiplied. The first term in eq.~(\ref{omega2}) $\omega_0^2 f_{2,0}$ has  maximal transcendentality five while $\omega_0$ has transcendentality one thus $f_{2,0}$ must have maximal transcendentality 3. Using similar arguments one can see that $f_{1,1}$ must have maximal transcendentality four and finally $f_{0,2}$ is of maximal transcendentality five.  
 The complexity of the functions increases  with maximal transcendentality~\footnote{The complexity of the function is not strictly defined and here we use ad hoc definition which includes both the transcendentality (the weight) and the depth of the harmonic sums which we define below. Higher transcendentality allows for larger depth, which evetually introduced functions of increasing complexity. },
  which is related to a  number and a sort  of nested summations used for defining a given function as we discuss in the next section. This also significantly increases a space of functions defining a functional basis for any ansatz.

\section{Recursive analytic solution using   roots of LO eigenvalue}\label{ch:recursive}

In the previous section we have discussed the complexity of the unknown functions building the NNLO eigenvalue. Each of those functions is a sum of two terms 
\begin{eqnarray}\label{hermit1}
f(z, \bar{z})= F(z)+F(\bar{z})
\end{eqnarray}
where $F(z)$ is the  single valued   function. A product of two such functions 
\begin{eqnarray}\label{hermit1}
f(z, \bar{z})g(z, \bar{z})= \left(F(z)+F(\bar{z})\right)\left(G(z)+G(\bar{z})\right)
\end{eqnarray}
is not reducible to one dimensional problem due to the cross term 
$F(z)G(\bar{z})$. 
In this section we propose a systematic iterative procedure of reducing the NLO expression to one dimensional problem at each iteration step. 
Firstly, we note that the expression in eq.~(\ref{omega2}) can be written as a polynomial of $\omega_0=a f_{0,0}$ as follows 
 
\begin{eqnarray}\label{omega20}
 \omega_2 
 =  a^3 \left( f^2_{0,0} f_{2,0}+f_{0,0}f_{1,1}+f_{0,0}f^2_{1,0}+
 f_{1,0}f_{0,1}+f_{0,2}\right)
\end{eqnarray}

In this representation we are  left with three unknown functions $f_{1,1}$, $f_{0,2}$ and $f_{2,0}$ in an explicit way. Three other functions $f_{0,0}$, $f_{0,1}$ and $f_{1,0}$ are known from the previous orders.

 The transcendentality arguments discussed in the previous section apply here as well so that the unknown function $f_{0,2}$ is the  most complicated function of maximal transcendentality five, the unknown function  $f_{1,1}$ is  of maximal transcendentality four and the unknown function  $f_{2,0}$ is  of maximal transcendentality three.
It is crucial to know the transcendentality of each function because 
it defines the number of free coefficients to be fixed in the
functional basis for each case.

The analytic continuation of the harmonic sums to the complex plane has been recently widely used to build the functional basis. The transcendentality five implies maximal weight of harmonic sums to be five, the transcendentality four implies maximal weight to be four etc. The number of free coefficients to be fixed is directly related to the maximal weight, at weight 3 there are 32 terms in the functional basis, at weight 4 there are 95 terms and finally at the weight 5 there are 288 free coefficients to be fixed. 

As it was already mentioned the analytic expression for the full functional dependence of $\omega_2$ on $z$ and $\bar{z}$ is not known. However, we do have analytic expressions for definite values of $\nu$ or $n$ in terms of the analytically continued harmonic sums calculated in    recent works N.~Gromov, F.~Levkovich-Maslyuk and G.~Sizov~\cite{GROMOV}, 
M.~Alfimov, N.~Gromov and G.~Sizov~\cite{GromovNonzero}
  and S.~Caron Huot and M.~Herraren~\cite{HUOT}. 
 Let us consider $n=0$, in this case $z$ and $\bar{z}$ are not independent anymore
 \begin{eqnarray}\label{zminusz}
 z+\bar{z}=-1
 \end{eqnarray} 
 and 
 then the NNLO eigenvalue can be written as 
 \begin{eqnarray}\label{F5}
 \omega_2(z, -1-z)= F_5(z)+F_5(-1-z)
 \end{eqnarray}
 where $F_5$ is the  single valued function calculated in Ref.~\cite{GROMOV}.
 Having the analytic expression for a particular limit of $n=0$  can be used  to find the full analytic form of $
 \omega_2$ in the following way. 
 
 Firstly, we realize that a rather simple separable form of eq.~(\ref{F5}) is a  result of pole decomposition of cross terms of type  $F(z) G(-1-z)$, which can be easily found using reflection identities of harmonic sums calculated by the authors in a series of publications~\cite{refl5,alter,refl4, refl2}.

Here we use the  fact that the harmonic sums are meromorphic functions with isolated poles located at negative integers.   The product of two harmonic sums $F(z) G(-1-z)$ have poles at  both, positive  and  negative integers. The reflection identities discussed in the next section separate this product into a sum of two meromorphic functions having poles at either negative integers or positive integers and zero. This pole separation plays a crucial role in our analysis and helps to restore  a sum of two pole separated functions (at $n=0$ or any other fixed $n$) into  products of mixed pole structure. The resulting expression is valid for any value of the conformal spin $n$.

  Solving the inverse problem of gathering together separable terms into cross product of different arguments is not an easy task and requires  some additional information, say at $n=1$ which was calculated by S.~Caron Huot and M.~Herraren~\cite{HUOT}. We will discuss this alternative approach in our further publications.

\begin{figure}[h!]
  \includegraphics[width=\linewidth]{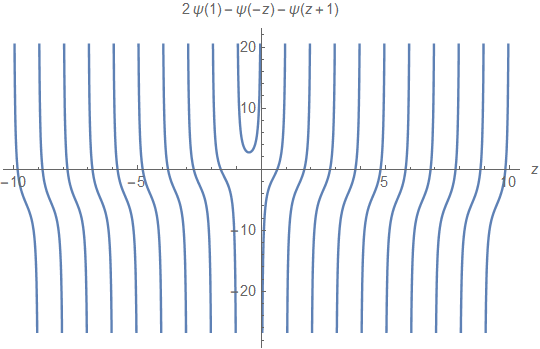}
  \caption{The figure shows zeros of $f_{0,0}$ for $n=0$, namely zeros of $2 \psi(1)-\psi(z+1)-\psi(\bar{z}+1)$ for $\bar{z}=-1-z$. The infinite number of zero roots $z_k$ for each $n$ allows to fix any finite set of free coefficients in the suggested ansatz.   }
  \label{fig:zerosn0}
\end{figure}

  We note that the problematic cross terms in eq.~(\ref{omega2}) come from  powers of $\omega_0$, which are proportional to (see eq.~(\ref{flo}))
  \begin{eqnarray}\label{psi}
  -\psi(z+1)-\psi(\bar{z}+1)+2 \psi(1)
  \end{eqnarray} 
where $\psi(z)= \frac{d \ln \Gamma(z)}{dz }$ is digamma function. 
The    function in eq.~(\ref{psi}) has infinite number of zero roots $z_k$ (here $k$ is an index labeling roots and  not directly related to conformal spin  $n$ ) consistent with eq.~(\ref{zminusz}), where it vanishes as shown in Figure \ref{fig:zerosn0}.

 At those roots the NNLO expression in eq.~(\ref{omega20}) reduces to 
\begin{eqnarray}\label{omega20zn}
 \left(\frac{\omega_2}{a^3}-f_{1,0}f_{0,1}\right)|_{z=z_k}=  f_{0,2}|_{z=z_k}
\end{eqnarray}
 Using this procedure we find values of $f_{0,2}$ at the zero roots of $f_{0,0}$, namely at $z=z_k$.  There are infinitely many of them so that we can fix any finite set of unknown coefficients provided we know the functional basis and it is finite.

 Note, that the function $f_{0,2}$ in eq.~(\ref{omega20zn}) has a separable pole structure   and thus can be directly compared to the known expression in eq.~(\ref{F5}) to fix the required 288 coefficients numerically~\footnote{The 288 terms of the functional basis include also constant terms like $\zeta(5)$ and $\zeta(3)\zeta(2)$, which are quite tricky for distinguishing numerically, but those can be unified into one constant reducing the number of the basis functions.}

After all free coefficients for $f_{0,2} $ are fixed and $f_{0,2}$ is known, one can subtract it from $\omega_2$, divide  by $f_{0,0}$
\begin{eqnarray}
 \frac{1}{f_{0,0}}\left(\frac{\omega_2}{a^3}-f_{1,0}f_{0,1}-f_{0,2}-f_{0,0}f^2_{1,0}\right)|_{z=z_k}=  f_{1,1}|_{z=z_k}
\end{eqnarray}
and then repeat the same procedure to find a simpler function $f_{1,1}$. 
At the last iteration we plug the known $f_{1,1}$ and  write 
\begin{eqnarray}
\frac{1}{f^2_{0,0}} \left(\frac{\omega_2}{a^3}-f_{1,0}f_{0,1}-f_{0,2}-f_{0,0}f^2_{1,0}-f_{0,0} f_{1,1}\right)|_{z=z_k}=  f_{2,0}|_{z=z_k}
\end{eqnarray}
and can finally find the last unknown function $f_{2,0}$. 

This way we show that it is possible to calculate the full functional form of $\omega_2$   by iterating it at the zero roots of $\omega_0$. 
The only possible issue related to this approach is that functions $f_{i,j}$ might  have the same roots as $\omega_0$, which is very unlikely based on the functions building the known NNLO eigenvalue for $n=0$ in eq.~(\ref{F5}). Suppose we do have some overlap of roots of $\omega_0$ and $f_{i,j}$, this can be resolved and cross checked by choosing another set of  roots as there are infinitely many of them. 

This procedure being conceptually simple is currently  difficult to implement due to small radius of convergence of integral representations of harmonic sums. The small radius of convergence   limits the grid and is problematic even for a very high precision calculations  because the harmonic sums are mostly slowly varying functions inside the radius of convergence. This leads to highly singular matrices for free coefficients, for which it is a computationally challenging problem to find  the  inverse  matrix.  We believe 
this technicality will be overcome in the nearest future.  

Note, that   the proposed iterative procedure allows to restore the full functional form of $\omega_2$ using only $n=0$ result by N.~Gromov, F.~Levkovich-Maslyuk and G.~Sizov~\cite{GROMOV}.

Other known cases   for $n \neq 0 $ and $\nu=0$ in  Ref.~\cite{Alfimov:2020obh} as well as for $n=1,2,..$ calculated by S.~Caron Huot and M.~Herraren~\cite{HUOT}, can be left  for cross checking the result.

\section{Comparison with known results in $N=4$ SYM}\label{comparison}

In an attempt to compare our ansatz in eq.~(\ref{omega2}) with the known 
results we analyze the $n=0$ case of the BFKL eigenvalue in $N=4$ SYM
calculated by N.~Gromov, F.~Levkovich-Maslyuk and G.~Sizov~\cite{GROMOV}. Their  result  is presented in terms of the harmonic sums $S_{a_1, a_2, ..., a_n} (z)$ analytically continued from  even integers values of the argument to the complex plane. 
The nested harmonic sums are defined~\cite{HS1,Vermaseren:1998uu,Blumlein:1998if,Remiddi:1999ew} as  nested summation for $n\in \mathbb{N}$~\footnote{Strictly speaking this definition holds for any $n\in \mathbb{N}$ only if $a_i>0$ because even and odd values of the argument $n$ should be treated separately if at least one of $a_i$ is negative as discussed below. }
\begin{eqnarray}\label{defS}
S_{a_1,a_2,...,a_k}(n)=  \sum_{n \geq i_1 \geq i_2 \geq ... \geq i_k \geq 1 }   \frac{\mathtt{sign}(a_1)^{i_1}}{i_1^{|a_1|}}... \frac{\mathtt{sign}(a_k)^{i_k}}{i_k^{|a_k|}}
\end{eqnarray}

We discuss   the harmonic sums with only real integer values of $a_i$, which build the alphabet of the possible negative and positive indices, which  uniquely  label $S_{a_1,a_2,...,a_k}(n)$.  
In eq.~(\ref{defS}) $k$ is  the depth and $w=\sum_{i=1}^{k}|a_i|$ is the weight of the harmonic sum $S_{a_1,a_2,...,a_k}(n)$.

 There are two different analytic continuations of
the harmonic sums~\cite{Velizhanin}   $a)$ the analytic
continuation from the even integer values of the argument and $b)$ the analytic continuation from the odd integer values to the complex plane.  Following the notation of Ref.~\cite{GROMOV} we use the analytic continuation of the harmonic sums from the even integer values of the argument.  The argument of the harmonic sums is a complex number 
\begin{eqnarray}
z=-\frac{1}{2}+\frac{ i \nu}{2} +\frac{|n|}{2}, \;\;\;
 \bar{z}=-\frac{1}{2}-\frac{ i \nu}{2} +\frac{|n|}{2},
\end{eqnarray}
where $\nu$ is continuous and real valued anomalous dimension~\footnote{By anomalous dimension $\nu$ we mean the parameter labeling $SL(2, \mathbb{C})$ representation of Lipatov spin chain. It is connected to the full dimension $\Delta$ of the twist-2 operators in $\mathit{N}=4 $ SYM as $\Delta=i \nu$. }  and $n$ is a conformal spin which takes   integer values.  
In our analysis we use the reflection identities for harmonic sums recently calculated by the authors up to weight of five~\cite{refl5,alter, refl4, refl2}.

The harmonic sums   are defined for positive integer argument
n and require an analytic continuation to the complex plane if one wishes to use
them as a general functional basis.
The reflection identities allow to decompose a product of two harmonic sums of different arguments $z$ and $-1-z$ into a sum of two sets of harmonic sums each of whom separately depends either on $z$ or $-1-z$. The reflection identity can be schematically written as follows
\begin{eqnarray}
S_{\{a\}}(z)S_{\{b\}}(-1-z)= S_{\{c\}}(z)+...+S_{\{d\}}(-1-z)+...
\end{eqnarray}
where  $\{a\}$,  $\{b\}$,  $\{c\}$ and  $\{d\}$ are sets of letters building the indices of the harmonic sums. There is fixed number of reflection identities at any given weight,for example,  there are $216$ irreducible reflection identities at weight $w=5$. All of the reflection identities up to weight of five were calculated by the authors in Ref.~\cite{refl5,alter, refl4, refl2}.

The simplest reflection identity at weight  $w=2$ reads
\begin{eqnarray}
S_{1}(z)S_{1}(-1-z)= S_{1,1}(z)+S_{1,1}(-1-z)+\frac{\pi^2}{3},
\end{eqnarray}
where $S_{1,1}(z)$ can be written as 
\begin{eqnarray}
S_{1,1}(z)= \frac{1}{2} \left(S_{1}(z)\right)^2+\frac{1}{2} S_{2}(z)  
\end{eqnarray}
using the  quasi shuffle identities of harmonic sums. 

The reflection identities are particularly useful in restoring the original expression from its pole decomposed form fro any specific value of the conformal spin.

For completeness  of our discussion we write the known LO and NLO BFKL eigenvalues given in eq.~(\ref{flo}) and eq.~(\ref{fnlo}) in terms of the harmonic sums analytically continued from even integer values of the argument to the complex plane as follows
\begin{eqnarray}\label{flos}
f^{LO}(z, \bar{z})= S_1 (z)+S_1 (\bar{z})
\end{eqnarray}
and 
\begin{eqnarray}\label{fnlos}
f^{NLO}(z, \bar{z})= \Phi(z+1)+\Phi(\bar{z}+1)-\frac{1}{2} f^{LO}(z, \bar{z}) \left(S_{-2}(z)+S_{-2}(\bar{z})\right) \;\;
\end{eqnarray}
where 
\begin{eqnarray}
\Phi(z+1)= 4 S_{1,-2}(z)-2 S_{-3}(z)  +2 S_3(z)+\frac{\pi^2}{3} S_1 (z).
\end{eqnarray}
 
The reflection identities can be used for pole separated decomposition of the term $S_1 (z) S_{-2} (\bar{z})$ in eq.~(\ref{fnlos}).

We use the reflection identities and  apply  our ansatz in eq.~(\ref{omega2}) to the result of Ref.~\cite{GROMOV}. Our analysis shows that  the most complicated part of the BFKL eigenvalue for arbitrary values of $\nu$ and $n$ takes  the following form 
\begin{eqnarray} \label{ansatzS}
F_3(z,\bar{z})& =& -128 \left( S_{1}(z) S_{1,-2,1}(\bar{z})+ S_{1}(\bar{z}) S_{1,-2,1}(z)\right)\\
&&
+256 \left( S_{1,1,-2,1}(z)+S_{1,1,-2,1}(\bar{z})\right)+ \textrm{simpler functions} \nonumber
\end{eqnarray} 

By \textit{simpler functions} we mean the harmonic sums of lower \textit{depth}~\footnote{The depth of harmonic sums is defined as a number of nested summations, which is equivalent to a number of letters in the index.} for any given weight, one can consider  the harmonic sums of the same weight five, but each having different depth, i.e. different complexity. The notion of "complexity" here is defined  \textit{ad hoc} and includes both the weight and the depth of the harmonic sums. Larger depth and higher weight imply higher complexity of a function corresponding to  the harmonic sum. In the perturbative expansion in $N=4$ SYM we deal with the harmonic sums of the same weight at any given order and thus higher complexity means only larger depth removing any ambiguity in the definition.

 For example, consider the function  $S_{1,1,-2,1}(z)$ has    \textit{depth} four, whereas $S_{2,-2,1}(z)$ has   depth three and 
$S_{-4,1}(z)$ has depth two. The term 
$S_{1}(z) S_{1,-2,1}(\bar{z})$ is  the most complex in the sense that it has the highest depth compared to any other function or any other product emerging in the final result. The depth $d$ is additive for the functions in the product in the sense of the adding a number of nested summations  minus one, i.e. adding $d-1$ for each term in the product. For example, the term $S_{1}(z) S_{1,-1,1,1}(\bar{z})$ would be more complex than any term in our result in eq.~(\ref{ansatzS}) 
  but according to our analysis 
such terms are absent as well as   $ S_{1}(z) S_{-2,1,1}(\bar{z})$, $ S_{1}(z) S_{1,-2,1}(\bar{z})$ and $ S_{1}(z) S_{1,1,-2}(\bar{z})$ terms.

For  zero conformal spin $n=0$  the variables $z$ and $\bar{z}$ are not independent anymore and related by $\bar{z}=-1-z$. In this case  eq.~(\ref{ansatzS}) reproduces the result of Ref.~\cite{GROMOV}~\footnote{See a useful representation of it in  eq.~(C.3) of the paper by S.~Caron Huot and M.~Herranen~\cite{HUOT}.}
\begin{eqnarray}
F_3(z,\bar{z})|_{n=0} & =& F_3(z) +F_3(\bar{z})= 
 -256 \left( S_{1,-2,1,1}(z)+S_{1,-2,1,1}(\bar{z})\right) \nonumber
 \\
 &&+ \textrm{simpler functions} 
\end{eqnarray}

The expression in eq.~(\ref{ansatzS}) corresponds to the term  $\omega_0 \; f_{1,1}$ in eq.~(\ref{omega2}). Note that the term
 $\omega^2_0 \; f_{0,2}$ in eq.~(\ref{omega2}) is absent in eq.~(\ref{ansatzS}) meaning that  $f_{0,2}=0$.

 For $n=1$  the relation between $z$ and $\bar{z}$ is slightly different and reads $\bar{z}=-z$, which shifts the argument of the harmonic sum by unity resulting into the "one over the argument" terms. In this case we have
 \begin{eqnarray}\label{F3N1our}
F_3(z,\bar{z})|_{n=1} & =& F_3(z) +F_3(\bar{z})=  \\
&&  -128  \frac{S_{-2,1,1}(z)}{z}+128 \frac{S_{1,-2,1}(z)}{z}  \nonumber
 \\
 &&-128  \frac{S_{-2,1,1}(\bar{z})}{\bar{z}}+128 \frac{S_{1,-2,1}(\bar{z})}{\bar{z}} \nonumber \\
 && + \textrm{simpler functions} \nonumber 
\end{eqnarray}

The corresponding term for $n=1$ calculated by Caron Huot and Herraren~\cite{HUOT} (see eq.~(C.5) of their paper)  reads
 \begin{eqnarray}\label{F3N1simon}
\tilde{F}_3(z,\bar{z})|_{n=1} & =& \tilde{F}_3(z) +\tilde{F}_3(\bar{z}) =  \\
&&
 -128  \frac{S_{-2,1,1}(z)}{z}+64  \frac{S_{1,-2,1}(z)}{z}
 -128  \frac{S_{1,1,-2}(z)}{z}  \nonumber   \\
&&
 -128  \frac{S_{-2,1,1}(\bar{z})}{\bar{z}}+64  \frac{S_{1,-2,1}(\bar{z})}{\bar{z}}
 -128  \frac{S_{1,1,-2}(\bar{z})}{\bar{z}}    \nonumber 
 \\
 && + \textrm{simpler functions} \nonumber 
\end{eqnarray} 
By direct comparison of two expressions in eq.~(\ref{F3N1our}) and eq.~(\ref{F3N1simon})  one can see that our result has a structure similar to that of S.~Caron Huot and M.~Herraren~\cite{HUOT}, with slightly different coefficients. 
We consider this result very encouraging especially in the light of the immense variety   of all possible functional forms available at weight $w=5$.

\section{Conclusion and Discussions}\label{}
In this paper we follow the  arguments of the paper by A.~Kotikov and L.~Lipatov~\cite{Kotikov:2001sc, Kotikov:2002ab}  and represent the BFKL equation as the Bethe-Salpeter equation for the scattering amplitude  with a leading simple pole in the plane of the complex angular momentum. We extend their analysis for next-to-leading~(NLO) to higher orders.   The corresponding all order equation  for the  eigenvalue is given in eq.~(\ref{Bethe}). We argue that this representation is more natural than the traditional Schr\"odinger-like equation in eq.~(\ref{scroedinger}) because it introduces a Hermitian separability of the BFKL eigenvalue in a natural way.  
 It is also  based on the analyticity, singularity and crossing symmetry of the scattering matrix and  being general should have the same form for both QCD and $N=4$ SYM. 
 The functions building it in QCD and $N=4$ SYM will be different, but the separability of the holomorphic and anti-holomorphic parts should be identical for both theories. 

A.~Kotikov and L.~Lipatov~\cite{Kotikov:2001sc, Kotikov:2002ab} showed that the Bethe-Salpeter approach introduces in a very natural way the hermitian separability to the BFKL equation at the NLO level allowing to factorize a complicated functions of two variables into a product of two much simpler one-variable functions.   We extend this analysis   to higher orders of the perturbation theory in eq.~(\ref{Bethe}). The next-to-next-to-leading~(NNLO) eigenvalue of the BFKL equation is of particular interest because the recent progress made using integrability techniques by  N.~Gromov, F.~Levkovich-Maslyuk and G.~Sizov~\cite{GROMOV}, 
M.~Alfimov, N.~Gromov and G.~Sizov~\cite{GromovNonzero}
  and S.~Caron Huot and M.~Herraren~\cite{HUOT}  for specific values of either conformal spin $n$ or anomalous dimension $\nu$ in $N=4$ SYM. The full analytic form of the NNLO  BFKL eigenvalue is still to be found.

We argue that the most complicated terms in the proposed ansazt for the NNLO eigenvalue is  a product of 
 functions of one complex variable. We also show that the unknown terms  are either hermitian separable on their own or hermitian separable functions multiplied by powers of the leading order~(LO) eigenvalue $\omega_0$ or other known functions.  In Section \ref{ch:recursive} we suggest an recursive approach for calculating  the unknown functions  based on  the zeros of $\omega_0$.  This approach should be also applicable to higher orders in the perturbative expansion.

\section{Acknowledgement}\label{}

We are indebted to Jochen Bartels, Victor Fadin,  Nikolay Gromov, Mikhail   Alfimov and Fedor Levkovich-Maslyuk  for inspiring discussions on the topic.


\end{document}